# Precision Cryptographic Calculation of the Observed Values of the Cosmological Constants $\Omega_\Lambda$ and $\Omega_m$ as a Manifestation of the Higgs State in the Extension Field $\mathbb{F}_{P_\alpha^2}$

Charles Kirkham Rhodes
Department of Physics, University of Illinois at Chicago,
Chicago, IL 60607-7059, *USA*

## ABSTRACT

The Higgs concept can be assigned a precise quantitative cosmic identity. Specifically demonstrated is the direct correspondence of the supersymmetric solution pair ($B_{Hh1}$ and $B_{Hh2}$) of the Higgs Congruence $B_{Higgs}^2 \equiv -1 \pmod{P_\alpha^2}$ in the extension field $\mathbb{F}_{P_\alpha^2}$ to the observed magnitudes of the cosmological constants $\Omega_\Lambda$ and $\Omega_m$. These results are in perfect agreement with the maximally preferred magnitudes of these quantities as experimentally determined ($0.712 < \Omega_\Lambda < 0.758$ and $0.242 < \Omega_m < 0.308$) by the concordance of measured ranges. The corresponding theoretical values found ($\Omega_\Lambda = 0.732688$ and $\Omega_m = 0.267312$) also satisfy exactly the condition $\Omega_\Lambda + \Omega_m = 1.0$ for perfect flatness, an outcome that is legislated by the concept of supersymmetry ($B_{Hh1} + B_{Hh2} = P_\alpha^2$) in $\mathbb{F}_{P_\alpha^2}$. The organizing principle underlying these findings stems from the identification of an observationally grounded modulus $P_\alpha$ and the subsequent construction of a modular counting system that correctly represents the physical entities. With $\phi(x)$ denoting the Euler totient function, it is shown that the Higgs Congruence can be reformulated to read $\phi(P_\alpha^2) + \phi(P_\alpha^2) \equiv B_{Higgs}^2 \equiv -1 \pmod{P_\alpha^2}$, an equivalence relation that, with mathematics left and physics right, provides a clear statement that the laws of modular counting in the two physically anchored finite fields $\mathbb{F}_{P_\alpha}$ and $\mathbb{F}_{P_\alpha^2}$ (1) govern the symmetry condition defining the Higgs state and (2) specify the exact values of $\Omega_\Lambda$ and $\Omega_m$ through the corresponding Higgs masses. Since previous work has established that the fine-structure constant $\alpha$ can be uniquely computed in $\mathbb{F}_{P_\alpha}$, in sharp accord with the best high-precision measurement (~370 ppt) of $\alpha$, the computation of $\Omega_\Lambda$ and $\Omega_m$ with the identical cryptographic apparatus demonstrates that a precise quantitative relationship exists between fundamental micro-scale couplings and the largest cosmic-scale entities. In fact, a simple physically motivated algorithm connects these vastly different physical quantities; in principle, given one member of the ($\alpha$, $\Omega_\Lambda$, and $\Omega_m$) triplet, the other two are uniquely determined. Furthermore, since an earlier analysis has shown that the symmetry defined by the Higgs state corresponds to the first supplementary law of Quadratic Reciprocity, that fundamental mathematical status is extended physically to $\Omega_\Lambda$ and $\Omega_m$, an act that lifts Quadratic Reciprocity into the cosmic realm. Since both fields $\mathbb{F}_{P_\alpha}$ and $\mathbb{F}_{P_\alpha^2}$ are fully defined by the prime $P_\alpha \equiv 1 \pmod 4$, the alliance of these quantitative findings relating $\alpha$, $\Omega_\Lambda$, and $\Omega_m$ provides additional confirming evidence for the correctness of its previously established physically based magnitude of $P_\alpha$. Ultimately, these results construct a coherent synthesis, in full conformance with observational data, that quantitatively relates the six physically intrinsic universal parameters α, G, h, c, $\Omega_\Lambda$, and $\Omega_m$.

# I.    INTRODUCTION

Prior studies [1,2] have demonstrated how a physically anchored cryptographic analysis [3-6] utilizing a finite field $\mathbb{F}_{P_\alpha}$, defined by an observationally determined prime modulus $P_\alpha = 1(\text{mod } 4)$, can precisely represent the particle mass scale and particle interactions. A key finding [7] was the successful computation of the fine-structure constant $\alpha$ in full agreement with the sharpest high-precision measurement, now standing at an exactitude of ~370 ppt. The basis of those results is a new organizing principle for the description of physical properties and interactions that is founded on underline{precise physically anchored modular counting} [7]. Although an integer of any magnitude can be expressed in $\mathbb{F}_{P_\alpha}$, it was found that $\mathbb{F}_{P_\alpha}$ was particularly suitable for the consideration of the masses of physical states whose values m fell below the Planck mass $m_{px} = \sqrt{\hbar c / G}$, the mass corresponding exactly [1,2,7] to the value of the modulus $P_\alpha$. Equivalently, for this class of systems, the normalized magnitudes of their corresponding mass numbers $B_m$ were bounded by the modulus of the field $P_\alpha$. Important examples of the results obtained [1,2,7] in $\mathbb{F}_{P_\alpha}$ were the predicted values of the masses of the electron and muon neutrinos, given respectively by $m_{\nu_e} = 0.8019\ meV$ and $m_{\nu_\mu} = 27.45\ meV$.

This work extends the previous analysis [7] by providing (A) a strong additional observational constraint governing the magnitude of the modulus $P_\alpha$ and (B) an evaluation of the efficacy of the extension field [3,5,8] $\mathbb{F}_{P_\alpha^2}$ for the representation of states whose masses considerably surpass the Planck value ($m_{px} \cong 21.7\mu g$). We will find that $\mathbb{F}_{P_\alpha^2}$ is indeed the natural voice for this genre of exceptionally massive states. Objects of high significance in this range of large masses fall into two classes. Specifically, they are (1) those possessing typical stellar magnitudes of ~$10^{33}$-$10^{34}$ g, since such heavy states may exhibit [9] the exceptionally energetic gamma-ray burst phenomenon [10-14], and (2) the cosmological entities $\Omega_\Lambda$ and $\Omega_m$, since they represent the most massive assemblies of energy and matter in the universe [15,16]. This study concentrates on the development of a precise cryptographic description of $\Omega_\Lambda$ and $\Omega_m$ and their connection to the concepts of supersymmetry and the Higgs state [1,2,7,17]. The chief aim is the calculation of their magnitudes with the underline{identical} cryptographic apparatus that has been previously developed [1,2,7] and the comparison of these results with the corresponding observational data. With respect to the earlier work on α [1,2,7] conducted in $\mathbb{F}_{P_\alpha}$, we emphasize that underline{no additional parameter or information of any kind has been introduced}. Overall, in concert with the earlier analysis [7] of the fine-structure constant α, this work culminates in a coherent consolidation of fundamental physical entities through the establishment of a set of precise quantitative statements that relate, in conformance with the corresponding observational data, the six intrinsic universal parameters α, G, h, c, $\Omega_\Lambda$, and $\Omega_m$ under the constraint of perfect flatness ($\Omega_\Lambda + \Omega_m = 1.0$).

## II.    DISCUSSION

### A.    Physical Conditions Establishing the Extension Field $\mathbb{F}_{P_\alpha^2}$

#### 1.    State Properties

The definition of a "critical state," as described in an earlier study [1], and the confirming correspondence of the calculated value [7] of the fine-structure constant α within ~150 ppt of the centroid of the best high-precision determination [18] of this parameter based on the measurement of g/2 illustrated in Fig.(1), have enabled the magnitude of the centrally important prime modulus $P_\alpha$ given in Table I to be established by a powerfully constraining interlocking grid composed of both physical data and independent mathematical requirements [7].  In summary, the cryptographic theoretical value of $\alpha$ is in accord with all past measurements of $\alpha$ including a recent determination based on a novel, but less precise method combining atomic interferometry with Bloch oscillations [19].  It will be shown below that the results described herein provide an additional independent quantitative constraint that specifically tests the magnitude of the modulus $P_\alpha$ and thereby further confirms the value [7] shown in Table I.

| Parameter | Integer | Prime Factors |
|---|---|---|
| $B_{Hl1}$ | 1464518162437886934175466847230415731421582430740004414836186 | 2, 3, 29, 37, 344353, 204901219, 14323448690905379, 22508563542484545478 0404199 |
| $B_{Hl2}$ | 5295067887317048385691240239639344309098972470130691907921115 | 5, 13, 31, 109, 677, 25717, 165235181, 3158418359, 5645180551, 470014220572174240309. |
| $k$ | 1806921683443759579805125284935052313575036953057098229031915 | 3, 5, 120461445562917305320341685662336820905002463537139881935461 |
| $-z$ | 4952664366311175740061581801934707726945517947813598093725385 | 3, 5, 719, 29537, 61069859, 250468882061, 240831741094489, 442085395250823 |
| $P_\alpha$ | 6759586049754935319866707086869760040520554900870696322757301 | $P_\alpha$ |
| $P_\alpha-1$ | 6759586049754935319866707086869760040520554900870696322757300 | $2^2$, $3^2$, $5^2$, 7, 11, 13, 17, 19, 23, 29, 31, 37, 41, 43, 47, 53, 59, 61, 67, 71, 73, 79, 83, 89, 97, 101, 103, 107, 109, 113, 127, 131, 137, 139, 149, 151 |
| $P_\alpha+1$ | 6759586049754935319866707086869760040520554900870696322757302 | 2, 8461, 45523, 83169760789807308284153, 1055047671275684092989413 25989 |
| $g_\alpha$ | 44419402346275330310319390315 0 (primitive root of $P_\alpha$) | 2, $3^2$, $5^2$, 7, 11, 17, 19, 31, 47, 53, 59, 61, 73, 79, 103, 109, 113, 131, 149 |
| $g_\beta^{-1}$ | 15217642950393595861435167215542 (primitive root of $P_\alpha$) | 2, 13, 23, 29, 37, 41, 43, 67, 71, 83, 89, 97, 101, 107, 127, 137, 139, 151 |
| $P_\alpha^{2}$ | 45692003564041530913706618865130769443723182401046215489977477767892955428481792811708758531308781179112886075267348804601 | $P_\alpha^2$ |
| $B_{Hh1}$ | 3347796059635390382881088364987808152071781494612682009037071198770067919259115346543183479220395326909020387514712622071 | $3^5$, 17, 23, 13221419329, 1093232496541527347545563496 19, 2558196946423985150753871157903541993, 952908891213839759379015119512941741 62369 |
| $B_{Hh2}$ | 1221404260440614053082553050014296129165140090643353348094040656912288750922267746516557505208838585220386568775263618253 0 | 2, 5, 46663, 190633, 22176136014869, 43515013978215459227, 41374940967071871512474984209536 5017, 34389534759271995742669674685321854760 5517 |

Table I: Compendium of Integers and Corresponding Prime Factors of Key Integers. Note that the $\gcd\left(g_\alpha, g_\beta^{-1}\right) = 2$ and $g_\alpha$ and $g_\beta^{-1}$ are both primitive roots of $P_\alpha$.

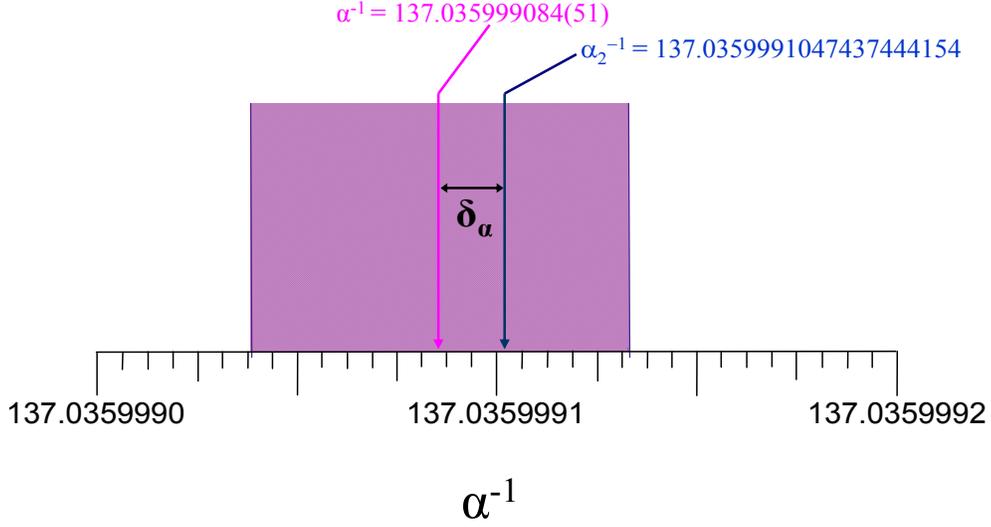

**Figure (1):** Comparison of the recently measured value [18] of the Fine-Structure Constant α with the formerly predicted [1] theoretical magnitude [7] $\alpha_2$. The magenta zone, representing the new high-precision (370ppt) data [18] based on a measurement of g/2 gives $\alpha^{-1} = 137.035999084$ (51). The cryptographic theoretical prediction yields $\alpha_2^{-1} = 4g_\beta^{-1} / g_\alpha = 137.035991047437444154$, a result that involves the two divisors $g_\beta^{-1}$ and $g_\alpha$ of $P_\alpha - 1$ that are the primitive roots of $P_\alpha$ presented in Table I. The theoretical procedure used to obtain this value was independently anchored by a set of interlocking constraints involving both physical observational data and mathematical stipulations [7]. The parameter $\delta_\alpha$, with $\delta_\alpha / \alpha^{-1} = 146$ ppt, quantifies the level of experimental/theoretical agreement with the high-precision g/2 data. Another experimental determination of α, founded on a new method that combines atomic interferometry with Bloch oscillations [19] fully overlaps the zone of $\alpha^{-1}$ illustrated and gives the specific value $\alpha_2^{-1} = 137.03599945$ (62). Accordingly, the theoretical magnitude of α, designated as $\alpha_2$ in the figure, is in agreement with all presently accepted observational values.

The system expressing the "critical" condition has a minimum energy [1] given by

$$E_{min} = 2(\hbar c^5 / G)^{1/2} = 2m_{px}c^2, \qquad (1)$$

a state that is physically represented by two noninteracting particles, each possessing a rest mass of $m_{px}$, that are at rest in their center-of-mass frame. This "critical state" also possesses a corresponding upper bound energy [1] in the same frame that represents the total mass/energy of the universe $M_u c^2$. The energy of this "cosmic bound" [1] is given by

$$E_{max} = P_\alpha m_{px} c^2 + \frac{m_{px} c^2}{P_\alpha} = m_{px} c^2 (P_\alpha + 1/P_\alpha), \tag{2}$$

a value that is directly proportional to the Planck mass $m_{px}$ and a multiplicative factor determined solely by the modulus $P_\alpha$. Although the analysis developed below requires both terms given by Eq. (2), we observe that the ratio of these terms is exactly $P_\alpha^2$, an integer whose magnitude exceeds $10^{121}$, thereby announcing an astonishing level of precision. Furthermore, with the energy unit of the system $E_0$ given by the extremely small second term ($m_{px} c^2 / P_\alpha$) in Eq. (2), we have

$$E_0 = (\hbar c^5 / G)^{1/2} / P_\alpha = m_{px} c^2 / P_\alpha = 1.8062 \times 10^{-33} eV, \tag{3}$$

a value whose uncertainty is determined by the CODATA [20] recommended experimental magnitude $(6.67428(67) \times 10^{-11} \mathrm{m^3 kg^{-1} s^{-2}})$ of the intrinsic universal constant G. We observe that $E_0$ is directly proportional to the Planck mass $m_{px}$, the utterly peerless physical choice for the universal definition of an energy scale. The availability of this unit enables the energies given by Eqs. (1) and (2) to be equivalently expressed in normalized dimensionless form respectively by

$$E_{min} / E_o = 2 P_\alpha \tag{4}$$

and

$$E_{max} / E_o = P_\alpha^2 + 1, \tag{5}$$

the pair of states illustrated to the left in the spectrum shown in Fig.(2). The difference in the normalized energies of these two states is, as indicated in Fig.(2),

$$\Delta = P_\alpha^2 - 2 P_\alpha + 1 = (P_\alpha - 1)^2 \equiv 1 (\mathrm{mod}\, P_\alpha). \tag{6}$$

Accordingly, the normalized energy $\Delta$, the enormous integer that gives the number of energy levels of the physical universe, also represents the unity residue class in the field $\mathbb{F}_{P_\alpha}$. We note additionally that the integer $(P_\alpha - 1)$ in Eq.(6) enjoys a dual mathematical significance, since the divisors of this number correspond to the subgroup orders of the cyclic group [3] of units $\mathbb{F}_{P_\alpha}^*$ associated with the field $\mathbb{F}_{P_\alpha}$.

Since Eq. (5) represents the total mass/energy of the universe [1], we have the mass number corresponding to the universe as

$$B_u = P_\alpha^2 + 1, \tag{7}$$

an even integer, thereby specifying fermi character [1,7,9], that is expressed by Eq.(7) in canonical p-adic form [21]. We observe additionally that the quantity $B_u$ represents the unity residue class in both of the fields $\mathbb{F}_{P_\alpha}$ and $\mathbb{F}_{P_\alpha^2}$, since the expression of $B_u$ in Eq.(7) simultaneously satisfies

$$B_u \equiv 1 (\mathrm{mod}\, P_\alpha) \tag{8}$$

and

$$B_u \equiv 1 (\mathrm{mod}\,\, P_\alpha^2). \tag{9}$$

It is useful to evaluate the mass number $B_{us}$ of the supersymmetric partner [1,2] of the universe in $\mathbb{F}_{P_\alpha}$. Since the mass parameters of all fermion/boson supersymmetric pairs ($B_f$ and $B_b$) in the cryptographic picture previously developed [1,2,7] satisfy the congruence

$$B_f + B_b \equiv 0 (\mathrm{mod}\, P_\alpha), \tag{10}$$

in which the subscripts f and b respectively denote fermi and bose species, we obtain

$$B_u + B_{us} = P_\alpha^2 + 1 + B_{us} \equiv 0 (\mathrm{mod}\, P_\alpha). \tag{11}$$

This statement gives the corresponding supersymmetric boson mass

$$B_{us} = m P_\alpha - 1 \tag{12}$$

for some even integer m, since, with $B_u$ even and $P_\alpha$ an odd prime, $B_{us}$ is perforce odd. Excluding negative values for $B_{us}$ in Eq.(12), $m \geq 2$, a condition that yields the minimum value

$$B_{us} = 2 P_\alpha - 1. \tag{13}$$

With reference to Eqs. (1) and (3), the mass number for $B_{us}$ is physically equal to

$$E_{min} - E_o = 2 m_{px} c^2 - (\hbar c^5 / G)^{1/2} / P_\alpha. \tag{14}$$

Equivalently, the mass number $B_{us}$ given by Eq.(13) represents a state that is composed of a pair of particles mutually at rest possessing the Planck mass $m_{px}$ shifted energetically downward by a term that (i) depends on the universal gravitational constant [22] G, (ii) is equal to the energy unit $E_o$ stated in Eq.(3), and (iii) corresponds to a decrease in the normalized energy given by its mass number of $\delta_g = E_o = 1$. Accordingly, the state described by $B_{us}$ can be regarded as the level $E_{min}$ after underline{gravitational renormalization} by $\delta_g = 1$, as shown in Fig.(2).

# Gravitational Renormalization of the Energy Structure of the Critical State

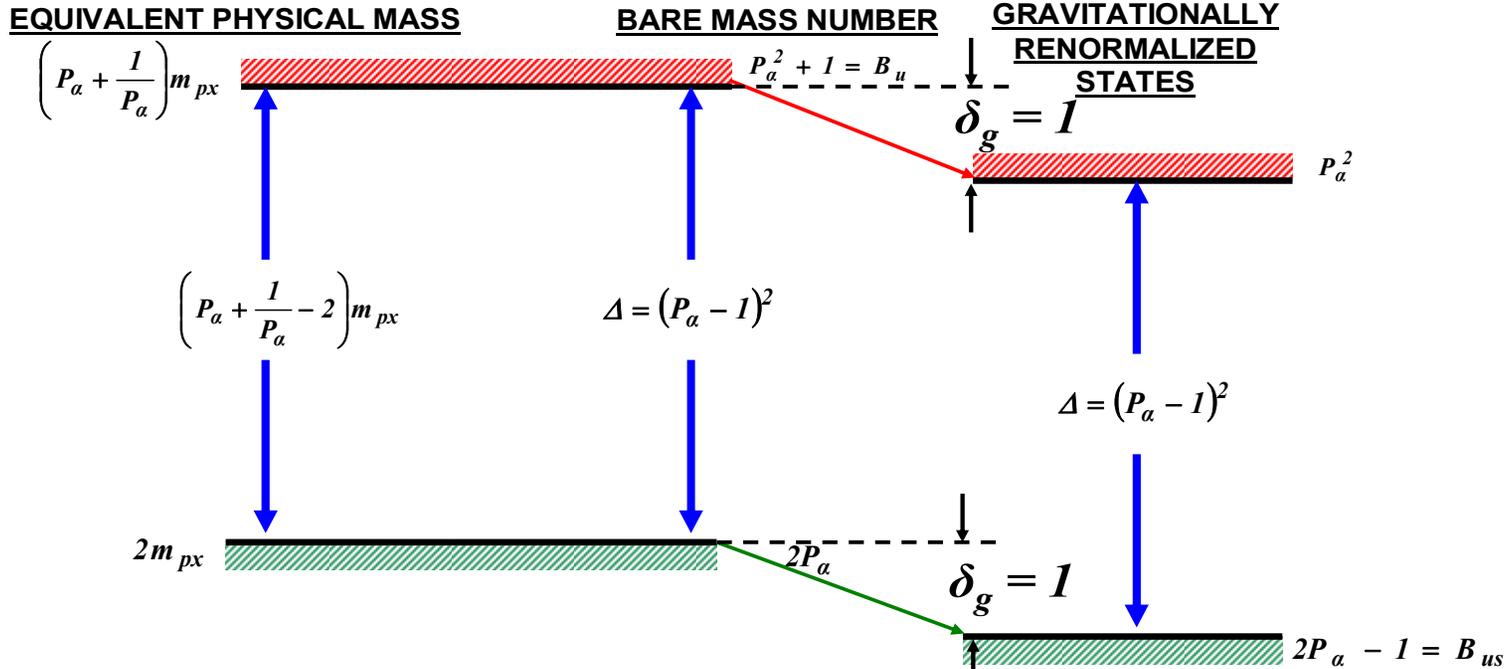

Fig (2): Spectrum of the "critical state" illustrating the ground level, with the bare mass number $2P_\alpha$ and equivalent mass $2m_{px}$, as well as the upper "cosmic bound" state corresponding to $B_u = P_\alpha^2 + I$, the bare mass number of the universe given by Eq.(5) with the physical bare mass $M_u$. The excitation energy $\Delta$ expressed by Eq.(6) separating these two states is given by $(P_\alpha - I)^2$. The gravitationally renormalized states are lowered by $\delta_g = I$, the energy unit $E_0$ of the mass scale and an amount physically equivalent to $m_{px} c^2/P_\alpha$ given by Eq.(3). This places the lower level at the mass number $B_{us} = 2 P_\alpha - I$, the value from Eq.(13) corresponding to the supersymmetric partner of $B_u$, and with an equivalent gravitational renormalization, locates the upper state at the mass number $P_\alpha^2$. A model of the gamma-burst phenomenon [9] has shown that all boson systems with mass numbers $B \le B_{us}$ are stable. The uniform gravitational renormalization of both levels by $\delta_g = I$ preserves the magnitude of $\Delta$. Both states are constructed to possess vanishing linear and angular momenta, the situation for a spin-zero boson at rest. Importantly, therefore, both underline{renormalized} states possess odd mass numbers, the parity of the integers associated with bose species and corresponding integer magnitudes for their spins; see text section II.A.2. With $2P_\alpha - I$ a known prime and $P_\alpha^2$ a prime power, both integers satisfy the requirements for respectively being moduli of the corresponding finite fields $\mathbb{F}_{P_\alpha^2}$ and $\mathbb{F}_{2P_\alpha - I}$

Therefore, the principle of supersymmetry expressed through Eqs.(10), (11), (12), and (13) establishes both the energy quantum of the system $E_o$ and the renormalization parameter $\delta_g$. It is also significant to recognize that the value of $E_o$ given in Eq.(3) is very close to the effective mass of fluctuations $(m_\phi \sim 10^{-33} eV)$ introduced in quintessence models [16]. This equivalence is physically quite natural, since these fluctuations should correspond to the minimal excitation of the system. Furthermore, for the physically anchored value of the prime $P_\alpha$, specific numerical test [7] discloses that the integer

$$P_\beta = 2P_\alpha - 1 \qquad (15)$$

is also a prime. This outcome invites the interpretation that the supersymmetric partner of the universe serves as the spatially fixed defining element of the cosmic realm, since particles represented by prime mass numbers perforce exhibit very robust confinement [1].

## 2. Gravitational Renormalization and State Parity

The two levels represented by Eqs. (4) and (5) shown in Fig. (2) are configured for vanishing angular momentum[1,2]. Hence, they represent states of bose character and should properly correspond to odd mass numbers [9]. However, the energy levels specified by the mass numbers given by Eqs.(4) and (5) are both even integers, the absolutely incorrect parity for this designation. The act of gravitational renormalization described above in relation to Eq.(14), physically demanded by the long range of this interaction, achieves the necessary correction. Specifically, the result given by Eq.(13) shows that the gravitational renormalization gives a <u>physical</u> value for $B_{us}$ that represents the minimum normalized energy level of the "critical state" given by Eq. (4) shifted downward exactly by unity. Accordingly, the energy of the "bare" state represented by Eq. (4) is gravitationally renormalized to the odd prime integer given by Eq. (15), the parity correctly corresponding to a boson. Mutatis mutandis, with the application of the same gravitational renormalization of unity to Eq. (4), the maximum energy level of the "critical state," likewise yields an odd mass number for the upper level denoted by $E_{max}$ in Eq.(5). Overall, we obtain from Eqs. (4) and (5), the modified levels

$$2P_\alpha \Rightarrow 2P_\alpha - 1 = P_\beta \qquad (16)$$

and

$$P_\alpha^2 + 1 \Rightarrow P_\alpha^2, \qquad (17)$$

as illustrated by the <u>gravitationally renormalized states</u> shown to the right in Fig. (2). This renormalization is key, since it gives both states odd mass numbers, hence, conferring bose character on these levels, an outcome essential for the designation of spin-zero configurations and a corresponding spatially isotropic universe. We observe that the uniform gravitational renormalization of these states also exactly preserves the magnitude of $\Delta$ given by Eq. (6); no observable effect on the spectrum of the system is caused by the renormalization.

Given the known properties [1,7] of the integer $P_\alpha - 1$, the arithmetic structure of $\Delta$ is quite exceptional in terms of its smoothness [23], factorability, and divisor structure. Specifically, $P_\alpha - 1$ is 151- smooth, possesses 36 distinct prime factors, and has $\sim 2.3 \times 10^{11}$ divisors [1,7], a figure close to the maximum possible for highly composite numbers (HCN) [24-27] with a magnitude of $\sim P_\alpha$. The latter property, which has been shown to be the direct consequence of (a) an optimization achieved by maximizing the complexity of the group structure of $\mathbb{F}_{P_\alpha}^*$ within the constraint of a bounded magnitude for the prime modulus $P_\alpha$ and (b) the simultaneous satisfaction of a physically based condition on the divisor structure of $P_\alpha - 1$ involving the Bézout identity [1,7], enables the set of divisors of this number [28,29] to represent in precisely encoded form an immense quantity of optimally organized information [17].

### B.    Mathematical Properties of the Extension Field $\mathbb{F}_{P_\alpha^2}$

The renormalized energy levels shown in Fig. (2) provide the physical basis for the use of the extension field $\mathbb{F}_{P_\alpha^2}$, since the energy of the upper state of the system given by Eq. (17) is represented by the mass number $P_\alpha^2$ and the power of a prime can validly serve as the modulus of an extension field [3,5,6,30,31] of the prime field $\mathbb{F}_{P_\alpha}$.

As demonstrated in prior analyses [1,2], the Higgs symmetry is expressed by the subgroup of order $\delta = 4$ in $\mathbb{F}_{P_\alpha}^*$. Accordingly, a supersymmetric pair of Higgs masses is immediately computed [1,2,17] by the solution of the Higgs congruence,

$$X^2 \equiv -1 (\bmod P_\alpha). \tag{18}$$

The demonstration that Eq. (18) defines integers $X$ of order four is readily achieved by squaring Eq. (18), an operation that yields

$$X^4 \equiv 1 (\bmod P_\alpha), \tag{19}$$

the canonical defining statement of such a number. The solution [1,2] of Eq. (18) gives a supersymmetric fermion/boson pair with both masses $> 10^{18} GeV$, but below the Planck mass $m_{px}$. Hence, the Higgs statement in $\mathbb{F}_{P_\alpha}$ yields two masses that fall in a relatively narrow range that approaches, but is slightly under the Planck scale. In this work, we designate the two known solutions of Eq.(18) in Table I as $B_{H\ell 1}$ and $B_{H\ell 2}$.

Since the definition of the Higgs symmetry [1,2,17] is expressed by the subgroup of order four, the concept of the Higgs congruence expressed by Eq.(18) can be directly transferred into and solved in the extension field $\mathbb{F}_{P_\alpha^2}$. By inspection, the congruence corresponding to Eq. (18) in $\mathbb{F}_{P_\alpha^2}$ is

$$X^2 \equiv -1 (\mathrm{mod}\ P_\alpha^2).$$ (20)

### C.    Solution of the Higgs Congruence in $\mathbb{F}_{P_\alpha^2}$

Tschebyscheff [32] showed that the quadratic congruence stated in Eq. (20) can be transformed into an equivalent linear congruence of the form

$$az \equiv b(\mathrm{mod}\ P_\alpha).$$ (21)

With the integers $a$ and $b$ known, the solution of Eq.(21) can be explicitly written by inspection as

$$z = a^{-1}b,$$ (22)

since

$$aa^{-1}b \equiv b(\mathrm{mod}\ P_\alpha)$$ (23)

from the general definition of an inverse element $a^{-1}$ through $aa^{-1} \equiv 1(\mathrm{mod}\ P_\alpha)$.

With this result, the two solutions of Eq. (20) can be written [32] as

$$B_{Hh1} = B_{Hl1} + zP_\alpha$$ (24)

and

$$B_{Hh2} = P_\alpha^2 - B_{Hh1},$$ (25)

where Eq.(25) follows from the expression of supersymmetric pairing in $\mathbb{F}_{P_\alpha^2}$ given by

$$B_{Hh1} + B_{Hh2} = P_\alpha^2.$$ (26)

We note that Eq. (26) simply generalizes the statement of supersymmetric pairing in $\mathbb{F}_{P_\alpha}$ given by Eq.(10); in both cases, the two solutions sum to the modulus of the corresponding field.

The governing equation [32] for the parameter $z$ in Eq. (24) is

$$\frac{B_{Hl1}^2}{P_\alpha} + 2B_{Hl1}z \equiv 0(\mathrm{mod}\ P_\alpha),$$ (27)

from which we obtain

$$z = -(2)^{-1}_{P_\alpha} (B_{Hh1})^{-1} \left[ \frac{B^2_{Hh1} + 1}{P_\alpha} \right].$$ (28)

From the symmetry of the Higgs state [1,2], we have

$$(B_{Hh1})^{-1} = B_{Hh2} \, ,$$ (29)

and the inverse integer

$$(2)^{-1}_{P_\alpha} = \frac{P_\alpha + 1}{2}$$ (30)

is readily calculated. The final results for $B_{Hh1}$ and $B_{Hh2}$ are presented in Table I. A direct substitution of these values of $B_{Hh1}$ and $B_{Hh2}$ into Eq. (20) explicitly verifies that the congruence is correctly satisfied.

The entries for $B_{Hh1}$ and $B_{Hh2}$ given in Table I, aside from their specific values, express an additional highly significant characteristic. It is the perforce innate and utterly stupendous precision of the cryptographic method. We observe from Eq. (20) the regency of one, while the magnitudes of $B_{Hh1}$ and $B_{Hh2}$ both exceed $10^{121}$; the exactitude of the result is greater than one part in $10^{121}$, the same scale encountered above in connection with Eq. (2). It follows that a hard epistemological limit is encountered [7], since no technology of measurement constructed and operated by man could ever achieve the precision necessary to achieve full verification of the theoretical values given for $B_{Hh1}$ and $B_{Hh2}$.

### D.    Super Seesaw Congruence in $\mathbb{F}_{P_\alpha^2}$

The original Higgs seesaw congruence [1,2,17]

$$g_\alpha g_\beta^{-1} = P_\alpha - 1 \equiv B^2_{Higgs} \equiv -1 (\bmod P_\alpha)$$ (31)

holds in $\mathbb{F}_{P_\alpha}$. In Eq. (31), the integers $g_\alpha$ and $g_\beta^{-1}$ given in Table I (a) are divisors of $P_\alpha - 1$, (b) are primitive roots of $P_\alpha$ that respectively denote prospective mass numbers [1,7] of the electron $\nu_e$ and muon $\nu_\mu$ neutrinos, and (c) have their magnitudes precisely pinned to the fine-structure constant [7] $\alpha$. The symbol $B_{Higgs}$ in Eq. (31) validly represents the full set of solutions to both Eqs. (18) and (20) specified by the integers $B_{H\ell1}$, $B_{H\ell2}$, $B_{Hh1}$, and $B_{Hh2}$ that are collectively presented in Table I. The validity of the solutions of Eq. (20) for Eq. (31) stems from the fact that any solution $(\bmod P_\alpha^2)$ is manifestly a solution $(\bmod P_\alpha)$. The correspondence of these solutions to a common residue class in $\mathbb{F}_{P_\alpha}$ is clear from Eq. (24), a point that is examined further in Section II.G.1. below.

We now demonstrate that a corresponding Higgs seesaw congruence exists in $\mathbb{F}_{P_\alpha^2}$. With reference to Eq. (31), the new Higgs congruence in $\mathbb{F}_{P_\alpha^2}$ can be readily seen to be

$$(P_\alpha + 1)g_\alpha g_\beta^{-1} = (P_\alpha + 1)(P_\alpha - 1) = P_\alpha^2 - 1 \equiv B_{Higgs}^2 \equiv -1 (\text{mod } P_\alpha^2), \qquad (32)$$

in which the symbol $B_{Higgs}$ denotes only the solutions $B_{Hh1}$ and $B_{Hh2}$ in Table I. Furthermore, since $P_\alpha + 1$ is an even composite integer whose prime factors appear in Table I, Eq. (32) can be equivalently expressed as

$$g_\gamma g_\delta^{-1} \equiv B_{Higgs}^2 \equiv -1 (\text{mod } P_\alpha^2), \qquad (33)$$

in which the new quantities $g_\gamma$ and $g_\delta^{-1}$ introduced are divisors of $P_\alpha^2 - 1$. This enables the new Super Seesaw Congruence in $\mathbb{F}_{P_\alpha^2}$, the counterpart of Eq. (31), to be stated in the form

$$g_\gamma g_\delta^{-1} = (P_\alpha + 1)g_\alpha g_\beta^{-1} = P_\alpha^2 - 1 \equiv B_{Higgs}^2 \equiv -1 (\text{mod } P_\alpha^2). \qquad (34)$$

### E.    Pattern of States in $\mathbb{F}_{P_\alpha^2}$

In parallel with the state representation [1,7] constructed in $\mathbb{F}_{P_\alpha}$, we can erect a corresponding description of the particle state pattern in $\mathbb{F}_{P_\alpha^2}$ and this basic state structure is shown in Fig. (3). The general morphology of the levels in $\mathbb{F}_{P_\alpha^2}$ is congruent to that [1,7] in $\mathbb{F}_{P_\alpha}$, but with the modification of $P_\alpha$ replaced by $P_\alpha^2$, there is an enormous magnification of the range of the mass scale.

The Super Seesaw Congruence given by Eq.(34) can be used to define states for $g_\gamma$ and $g_\delta^{-1}$ in Fig. (3). We assume that $g_\alpha$ and $g_\beta^{-1}$ are divisors of $P_\alpha - 1$ as given by Eq. (31). In parallel with the state pattern developed in $\mathbb{F}_{P_\alpha}$ [1,2], under the condition that $g_\gamma$ and $g_\delta^{-1}$ are divisors of $P_\alpha^2 - 1$, and with $g_\gamma$ corresponding to particle *P*, we have the following correspondences [2], as presented in Fig. (3). Specifically, they mirror the structure previously developed [1,2,7] in $\mathbb{F}_{P_\alpha}$ and are

$$P_{ss} = P_\alpha^2 - g_\gamma = g_{\delta,} \qquad (35)$$

$$(P_{in})_{ss} = (P_\alpha^2 - 1) / g_\gamma = g_\delta^{-1}, \tag{36}$$

and

$$P_{in} = g_\gamma^{-1} = P_\alpha^2 - g_\delta^{-1}. \tag{37}$$

By definition, it follows perforce that

$$g_\gamma g_\gamma^{-1} \equiv 1 (\mathrm{mod}\, P_\alpha^2) \tag{38}$$

and

$$g_\delta g_\delta^{-1} \equiv 1 (\mathrm{mod}\, P_\alpha^2). \tag{39}$$

### F.    Group Structure in $\mathbb{F}_{P_\alpha^2}$

Euler's totient function $\phi(n)$ generally [33] yields the magnitude of the multiplicative group of integers modulo $n$. It also gives the number of integers greater than zero and less than or equal to $n$ that are relatively prime to $n$. Specifically, for $n = P_\alpha^2$ we have [34]

$$\phi(P_\alpha^2) = P_\alpha(P_\alpha - 1) = P_\alpha(g_\alpha g_\beta^{-1}) = P_\alpha^2 - P_\alpha \equiv -P_\alpha(\mathrm{mod}\, P_\alpha^2). \tag{40}$$

Hence, in comparison with the field $\mathbb{F}_{P_\alpha}$, that possesses the corresponding totient function

$$\phi(P_\alpha) = P_\alpha - 1 = g_\alpha g_\beta^{-1} \equiv -1 (\mathrm{mod}\, P_\alpha), \tag{41}$$

the structure of the subgroup orders is uniformly expanded by a factor of $P_\alpha$, an outcome that parallels the magnification of the mass scale illustrated in Fig.(3). Furthermore, the reformulation of Eqs. (31) and (34) with the corresponding totient functions given by Eqs. (40) and (41) yields the comparative relations

$$\phi(P_\alpha) \equiv B_{Higgs}^2 \equiv -1 (\mathrm{mod}\, P_\alpha) \tag{42}$$

in $\mathbb{F}_{P_\alpha}$ and

$$\phi(P_\alpha^2) + \phi(P_\alpha) \equiv B_{Higgs}^2 \equiv -1 (\mathrm{mod}\, P_\alpha^2) \tag{43}$$

in $\mathbb{F}_{P_\alpha^2}$.

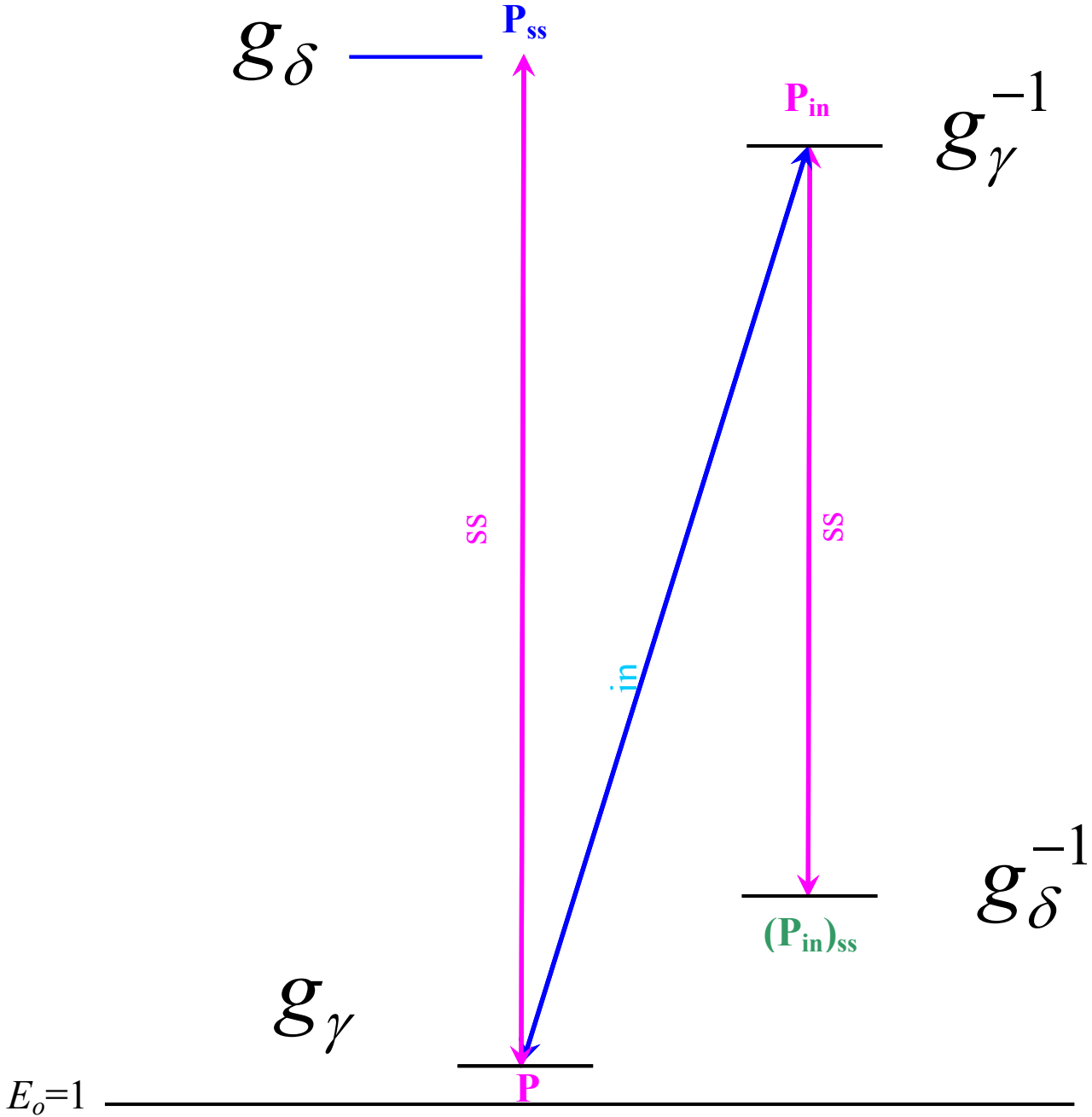

Fig. (3): Illustration of the general state pattern for the mass scale octet [1,2,7] in $\mathbb{F}_{P_\alpha^2}$. For particle $P$, the corresponding supersymmetric $P_{ss}$ and inverse $P_{in}$ states are shown. The quartet of levels represented is doubled to yield an octet by the existence of a corresponding set of four anti-particles with mass values degenerate to the states shown. The conditions of supersymmetry governing the respective mass numbers, illustrated for the Higgs states by Eq.(26), are $P + P_{ss} = P_\alpha^2$ and $(P_{in})_{ss} + P_{in} = P_\alpha^2$; the two mass values sum to the modulus. The energy unit $E_o$, given by Eq.(3), specifies the value of unity. The states $g_\gamma$, $g_\delta$, $g_\gamma^{-1}$, $g_\delta^{-1}$ following from Eq. (35), (36), and (37) are shown. Since $g_\gamma \mid P_\alpha^2 - 1$, it follows [1,2] that $g_\delta^{-1} = (P_\alpha^2 - 1)/g_\gamma$; the result demonstrated in $\mathbb{F}_{P_\alpha}$ carries through in $\mathbb{F}_{P_\alpha^2}$.

The two congruences expressed by Eq. (42) and (43) with the Euler totient functions relate the mathematical characteristics of the multiplicative groups directly to the physical definition of the Higgs symmetry [1,2]. The left hand sides of Eqs. (42) and (43) are composed purely of arithmetic functions [33,34], while the corresponding right hand sides express the physical concepts of mass and space [1] through the Higgs state. Accordingly, with mathematics on the left and physics on the right, these equivalence relations demonstrate that the laws of modular counting in the physically anchored finite fields $\mathbb{F}_{P_\alpha}$ and $\mathbb{F}_{P_\alpha^2}$ (a) govern the symmetry condition defining the Higgs state and (b) specify the exact values of the Higgs masses. The comparison of Eqs.(42) and (43) reveals an additional exceptional property. We observe that these two statements are linked by the term $\phi(P_\alpha)$, the value of which from Eq. (41) is $g_\alpha g_\beta^{-1}$, the integer representing respectively the product of the mass numbers of the electron $\nu_e$ and muon $\nu_\mu$ neutrinos. Accordingly, Eq. (43) connects precisely the lightest massive systems known, the $\nu_e$ and $\nu_\mu$ neutrinos, to the mass of the superlatively heavy Higgs states $B_{Hh1}$ and $B_{Hh2}$ presented in Table I, whose huge magnitudes are a factor of $\sim 10^{92}$ greater. This vast range points to a theoretical coherence that deeply penetrates a new realm of scale, since it is by far the largest span over which sharp quantitative relationships have been expressed for the magnitudes of any physical qualities.

### G. Properties of the Higgs Solutions

#### 1. Residue Class Membership in $\mathbb{F}_{P_\alpha}$

The specific solution of Eq. (20) for the mass number $B_{Hh1}$ of the Higgs system is stated by Eq.(24) with the parameter z given by Eq.(28). Since

$$B_{Hh1} - B_{H\ell1} = zP_\alpha \equiv 0 (\text{mod } P_\alpha), \tag{44}$$

the integers $B_{Hh1}$ and $B_{H\ell1}$ are representatives of the same residue class in $\mathbb{F}_{P_\alpha}$; basically, the Higgs symmetry specifies a single residue class. Accordingly, the heavy Higgs masses $B_{Hh1}$ and $B_{Hh2}$ possess the identical connection to the law of Quadratic Reciprocity [35] that has been previously established [17] for $B_{H\ell1}$ and $B_{H\ell2}$.

#### 2. Connection of Higgs Solutions in $\mathbb{F}_{P_\alpha}$ and $\mathbb{F}_{P_\alpha^2}$

The full explicit solution of Eq.(20) for $B_{Hh1}$, derived by combination of Eqs. (24), (28), (29), and (30), can be written as

$$B_{Hh1} = B_{H\ell1} + zP_\alpha = B_{H\ell1} - \left(\frac{P_\alpha + 1}{2}\right) B_{H\ell2} \left(B_{H\ell1}^2 + 1\right), \tag{45}$$

a result that illustrates another close connection between the solutions of Eqs. (18) and (20). The former solved in $\mathbb{F}_{P_\alpha}$, along with modulus $P_\alpha$, legislates the latter in $\mathbb{F}_{P_\alpha^2}$. However, all solutions fundamentally depend <u>only</u> on the prime modulus $P_\alpha$ and the order $\delta = 4$ of the subgroup that expresses the Higgs symmetry. Finally, since the magnitude of the first term $B_{H\ell 1}$ in Eq. (45) is very small, with

$$B_{Hh1} / B_{H\ell 1} \sim 10^{61} \tag{46}$$

as shown in Table I, we can imagine $B_{H\ell 1}$ as an echo of the residue class of the solution in $\mathbb{F}_{P_\alpha}$, the sotto voce that announces the solution for $B_{Hh1}$ in $\mathbb{F}_{P_\alpha^2}$.

The relative parities of the Higgs solutions $B_{H\ell 1}$ and $B_{Hh1}$ respectively found in $\mathbb{F}_{P_\alpha}$ and $\mathbb{F}_{P_\alpha^2}$ are organized by Eq. (44), and with $z$ an odd integer, as shown in Table I, they are opposite. This outcome yields heavier bosons than fermions in both fields.

## III.    INTERPRETATION OF THE HIGGS MASS NUMBERS $B_{Hh1}$ AND $B_{Hh2}$

The interpretations of the two Higgs states described by $B_{Hh1}$ and $B_{Hh2}$ in Table I are unambiguous, since only two physical entities are known to exist at the mass scale [15,16] they represent. Consider the following identifications that are computed with respect to the mass number of the universe $B_u$ given by Eq. (7). Specifically, we compute the ratio

$$\Omega_\Lambda = B_{Hh1} / B_u = B_{Hh1} / P_\alpha^2 + 1 \tag{47}$$

and the corresponding quantity

$$\Omega_m = B_{Hh2} / B_u = B_{Hh2} / P_\alpha^2 + 1. \tag{48}$$

However, since supersymmetry requires

$$B_{Hh1} + B_{Hh2} = P_\alpha^2, \tag{49}$$

we obtain the sum

$$\Omega_\Lambda + \Omega_m = P_\alpha^2 / P_\alpha^2 + 1, \tag{50}$$

whose value is unity to within one part in $\sim 10^{121}$, a level of deviation that is fundamentally refractory to measurement. However, the replacement in Eqs. (47), (48), and (50) of the <u>bare</u> mass $B_u$ with the corresponding <u>renormalized value</u> of $B_u$-1 from Eq. (17) converts these statements to

$$\Omega_\Lambda = B_{Hh1} / P_\alpha^2 = 0.732687 \,, \tag{51}$$

$$\Omega_m = B_{Hh2} / P_\alpha^2 = 0.267312 \,, \tag{52}$$

which yields <u>precisely unity</u> for the sum

$$\Omega_\Lambda + \Omega_m = P_\alpha^2 / P_\alpha^2 = 1.0 \,. \tag{53}$$

Hence, the concept of supersymmetry in $\mathbb{F}_{P_\alpha^2}$ expressed by Eq. (49) is equivalent to perfect flatness of the universe and the comparison of Eqs. (50) and (53) both resounds as a noble cosmic legato the importance of unity expressed by Eq. (20) and underlines the utterly fundamental need for the gravitational renormalization detailed in Fig. (2).

The computed values for $\Omega_\Lambda$ and $\Omega_m$ given by Eqs. (51) and (52) can be compared with two sets of measured values [36,37] that were obtained by fully independent procedures. One method [36] utilizes x-ray measurements from *Chandra* while the second approach [37] involves a geometric test (Alcock-Paczyski) associated with bound galactic pairs. As illustrated in Figs. (4) and (5), these comparisons demonstrate full agreement with the measured ranges and reveal that the natural correspondences are the identification of $B_{Hh1}$ with the Cosmological Constant $\Omega_\Lambda$ and the association of $B_{Hh2}$ with $\Omega_m$, the parameter that describes the matter in the universe. The observed and computed values of $\Omega_\Lambda$ and $\Omega_m$ comprehensively favor the existence of a flat universe, the outcome theoretically legislated by the statement of supersymmetry in Eq. (49). Hence, supersymmetry and flatness become identical statements that are respectively expressed in Eqs. (10) and (53).

## IV.    DIRECT CONNECTION OF α TO $\Omega_\Lambda$ AND $\Omega_m$

As developed above, the new modality of calculation that yielded the result for α illustrated in Fig. (1) has been successfully extended to the computation of the observed values [36,37] of the Cosmological Constant $\Omega_\Lambda$ and the associated cosmic parameter $\Omega_m$. Importantly, this computational extension was achieved <u>without the incorporation of additional information or parameters</u> of any kind. Therefore, it immediately follows that there exists perforce an algorithm that produces the observed magnitudes for $\Omega_\Lambda$ and $\Omega_m$ directly from the precise theoretical value of α. As shown below, this algorithm is direct, elementary, and physically based. The existence of this relationship is profoundly significant; it signals that a measurement of any one of the triad (α, $\Omega_\Lambda$, $\Omega_m$), in principle, is equivalent to the measurement of the remaining pair. Hence, an experimental determination of α in a deep cave produces a view of the heavens.

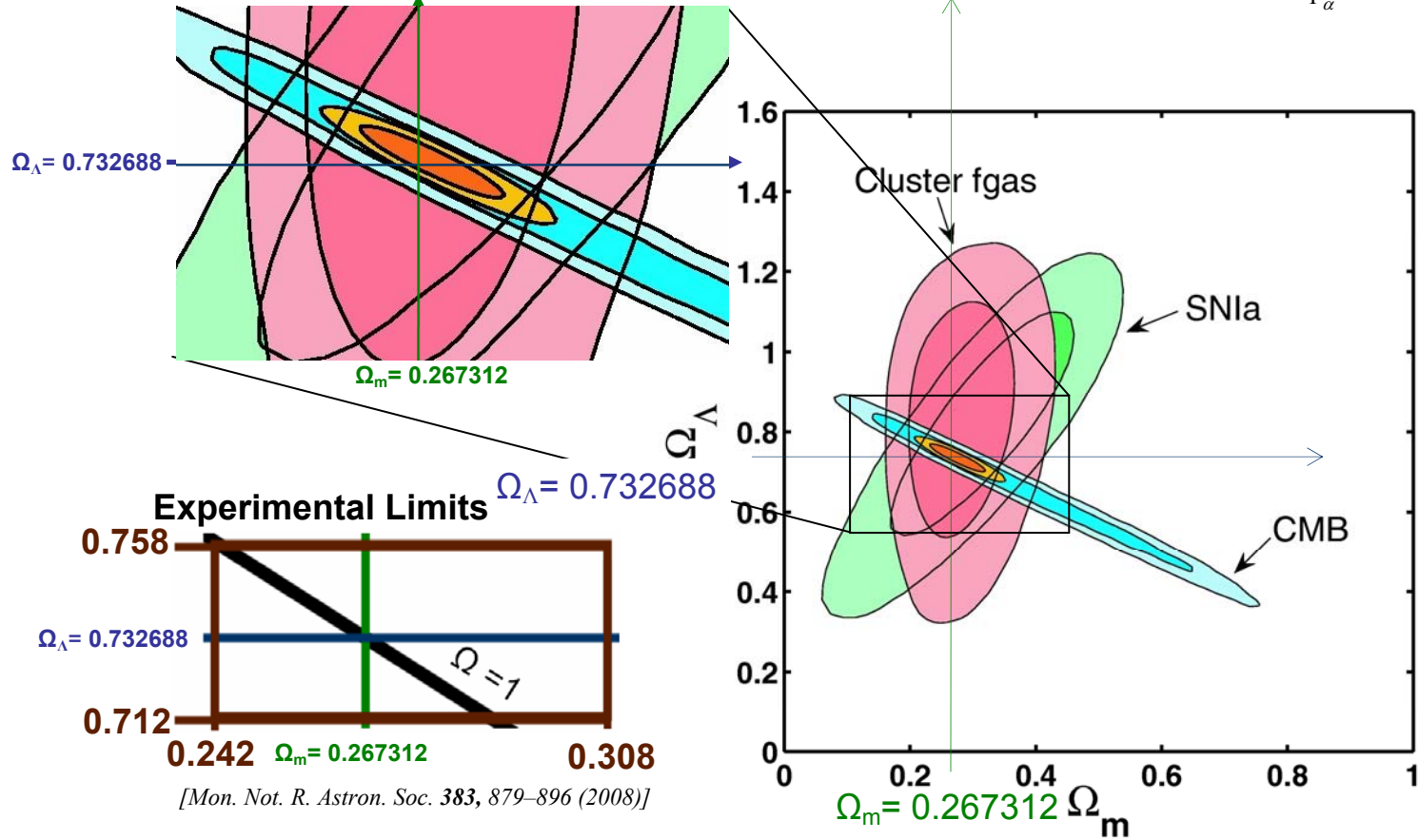

**Cryptographic Computation of $\Omega_\Lambda$ and $\Omega_m$ as Super Higgs States in $\mathbb{F}_{p_\alpha^2}$**

$\Omega_\Lambda$ = 0.732688

$\Omega_m$ = 0.267312

$\Omega_\Lambda$ = 0.732688

**Experimental Limits**

$\Omega_\Lambda$ = 0.732688

$\Omega$ =1

0.758

0.712

0.242   $\Omega_m$ = 0.267312   0.308

*[Mon. Not. R. Astron. Soc. **383**, 879–896 (2008)]*

$\Omega_\Lambda$

Cluster fgas

SNIa

CMB

$\Omega_m$ = 0.267312   $\Omega_m$

Fig. (4): Comparison of the computed values of $\Omega_\Lambda$ and $\Omega_m$ from the Super Higgs Congruence in $\mathbb{F}_{p_\alpha^2}$ with the assembly of correlated data restricting the ranges of $\Omega_\Lambda$ and $\Omega_m$. The concept of supersymmetry legislates the condition of $B_{Hh1} + B_{Hh2} = P_\alpha^2$, a statement equivalent to perfect flatness given by $\Omega_\Lambda + \Omega_m = 1.0$. The inset shown at the upper left details the central zone illustrating the agreement between the calculated and experimental values. The theoretical values of $\Omega_\Lambda$ and $\Omega_m$ are compared directly with the experimental limits at 68% confidence ($0.712 < \Omega_\Lambda < 0.758$, $0.242 < \Omega_m < 0.308$) in the box placed in the lower left panel. The flat universe $\Omega =1$ contour is shown for reference. The figure is adapted from Ref. [36] and used with permission.

# Cryptographic Computation of $\Omega_\Lambda$ and $\Omega_m$ as Super Higgs States in $\mathbb{F}_{P_\alpha^2}$

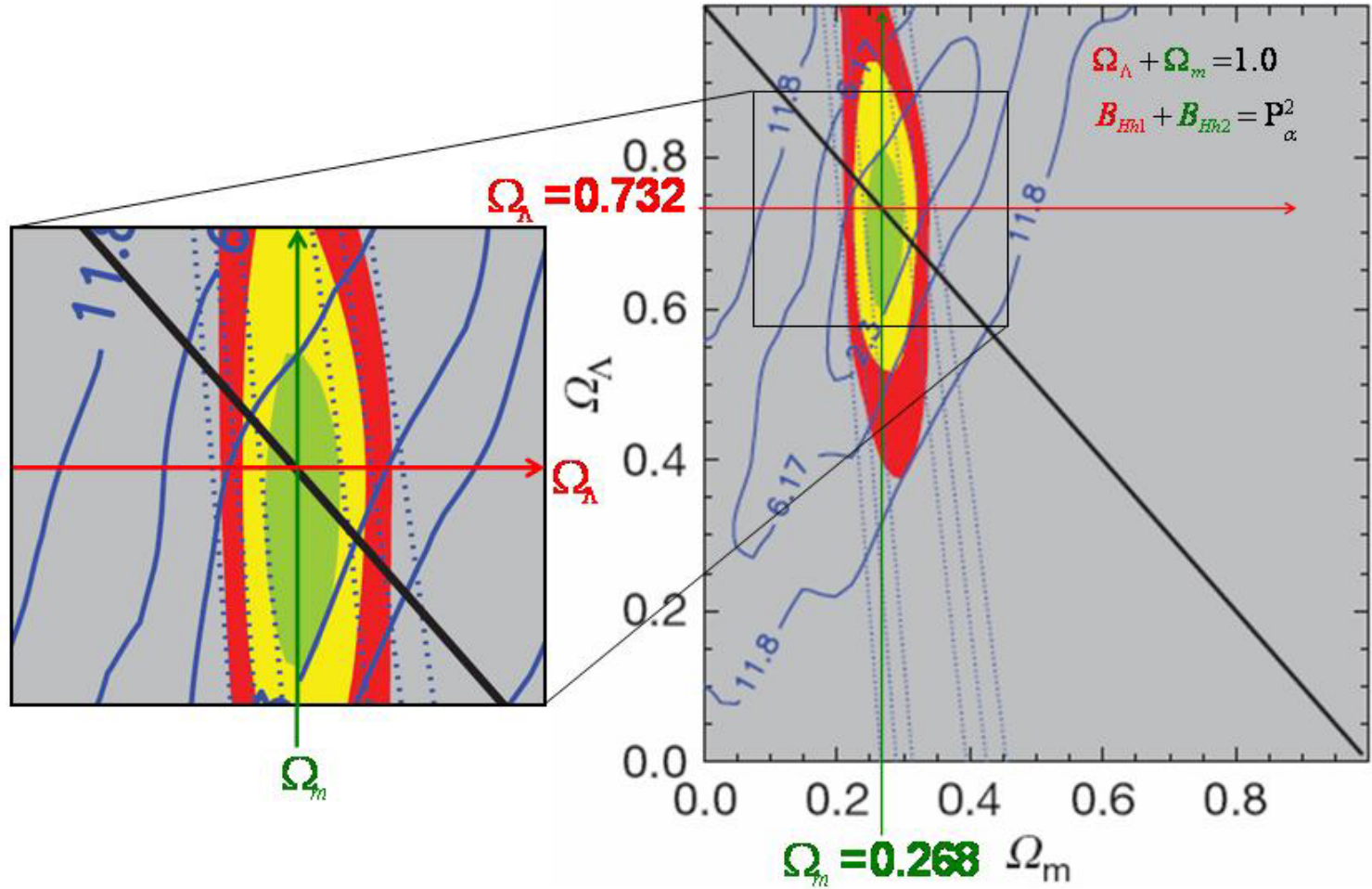

Fig. (5): Comparison of the computed values of $\Omega_\Lambda$ and $\Omega_m$ from the Super Higgs Congruence in $\mathbb{F}_{P_\alpha^2}$ with the assembly of correlated data restricting the ranges of $\Omega_\Lambda$ and $\Omega_m$ derived from a geometric measure based on bound galactic pairs [37]. Perfect agreement of the computed values with the data is manifest. The concept of supersymmetry legislates the condition of $X_1 + X_2 = P_\alpha^2$, a statement equivalent to perfect flatness given by $\Omega_\Lambda + \Omega_m = 1.0$, the condition specified by the black diagonal line. The figure is adapted from Ref. [37] and used with permission.

We now demonstrate the simplicity of the calculation that relates $\alpha$ to $\Omega_\Lambda$ and $\Omega_m$. Since $\alpha$ is theoretically composed as a rational number of the form

$$\alpha = \frac{g_\alpha}{4 g_\beta^{-1}} \tag{54}$$

with $g_\alpha$ and $g_\beta^{-1}$ known integers [1,7], we form the product

$$g_\alpha g_\beta^{-1} = P_\alpha - 1 = \phi(P_\alpha) \,, \tag{55}$$

obtaining the totient function of $P_\alpha$ given by Eq. (41). The subsequent addition of unity generates the quantity

$$P_\alpha = g_\alpha g_\beta^{-1} + 1 \,, \tag{56}$$

a step ordained by supersymmetry [1,7,17] yielding the modulus $P_\alpha$ given in Table I that represents the Planck mass $m_{px}$ and is a prime congruent 1(mod4) by direct test. We then square $P_\alpha$, employ the result $P_\alpha^2$ as a cosmic-scale modulus, and solve by a known procedure [32] the quadratic congruence

$$X^2 \equiv -1 \bmod (P_\alpha^2) \tag{57}$$

that mirrors Eq. (20) and specifies the Higgs symmetry [1,7,17]. This gives two solutions, $X_1$ and $X_2$, that, on rigorous mathematical grounds [32], obey the exact condition

$$X_1 + X_2 = P_\alpha^2 \,. \tag{58}$$

In parallel with Eqs. (51) and (52), normalization of Eq. (58) by $P_\alpha^2$ constructs the physical correspondences

$$\Omega_\Lambda = X_1 / P_\alpha^2 \,, \tag{59}$$

and

$$\Omega_m = X_2 / P_\alpha^2 \,. \tag{60}$$

Subsequently, from Eq. (58), the condition

$$\Omega_\Lambda + \Omega_m = 1.0 \tag{61}$$

follows, the statement of a perfectly flat universe [36,37].

This physically motivated mathematical outcome demonstrates the existence of an <u>exact corresponding physical connection</u> between these three profoundly fundamental entities, $\alpha$, $\Omega_\Lambda$, and $\Omega_m$, with $\alpha$ representing fundamental micro-scale couplings and $\Omega_\Lambda$ and $\Omega_m$ denoting the cosmic-realm. Furthermore, with the time dependence of $\alpha$ experimentally ruled out [38,39], this triple conjunction likewise legislates a corresponding temporal independence for both $\Omega_\Lambda$ and $\Omega_m$.

## V.   CONCLUSIONS

The solution of the Higgs Congruence $B_{Higgs}^2 \equiv -1 (\mathrm{mod}\, P_\alpha^2)$ in the extension field $\mathbb{F}_{P_\alpha^2}$ yields the cosmic parameters $\Omega_\Lambda$ and $\Omega_m$ in full accord with their experimental determinations. These solutions satisfy exactly the relation $\Omega_\Lambda + \Omega_m = 1.0$, the condition of perfect flatness whose fundamental basis is supersymmetry. The recasting of the Super Seesaw Congruence given by Eq. (34) into the form

$$\phi(P_\alpha^2) + \phi(P_\alpha) \equiv g_\gamma g_\delta^{-1} \equiv (P_\alpha + 1) g_\alpha g_\beta^{-1} \equiv B_{Higgs}^2 (\mathrm{mod}\, P_\alpha^2), \tag{62}$$

that explicitly involves the Euler totient function $\phi(x)$, demonstrates that the laws of modular counting in both $\mathbb{F}_{P_\alpha}$ and $\mathbb{F}_{P_\alpha^2}$ simultaneously (a) legislate the symmetry condition defining the Higgs state and (b) determine the precise values of $\Omega_\Lambda$ and $\Omega_m$. A <u>quantitative cosmic identity</u> is thereby conferred upon the Higgs concept; the Higgs boson is $\Omega_\Lambda$.

Finally, since the previous study [7] giving the theoretical value of the fine-structure constant $\alpha$, shown in Fig.(1) as the ratio

$$\alpha_2^{-1} = 4 g_\beta^{-1} / g_\alpha = 137.035991047437444154, \tag{63}$$

that involves the two primitive roots $g_\alpha$ and $g_\beta^{-1}$ of $P_\alpha$ shown in Table I, is a rigorous test of the magnitudes and orders of the <u>divisors</u> of $P_\alpha - 1$, while the solution of Eq. (20) for $\Omega_\Lambda$ and $\Omega_m$ is a completely independent probe of the <u>magnitude</u> of $P_\alpha$, the confluence of these quantitative findings interlocking the six intrinsic universal parameters $\alpha$, G, h, c, $\Omega_\Lambda$, and $\Omega_m$ expressed comprehensively through Eqs. (1), (2), (3), (11), (14), (17), (20), (47), (48), (53), and (63), that stand uniformly in conformance with the corresponding observational data, provides both a coherent theoretical synthesis of these fundamental entities and strong physically anchored evidence for the unique status of the value selected for the prime modulus $P_\alpha$ on which the numerical findings rest.

<u>Acknowledgements</u>

The author thanks John C. McCorkindale for his effort on the modular computations, Steven W. Allen for his valuable advice on the status of the observational data, and James W. Longworth for many informative discussions.

REFERENCES

1.  Dai, Y., Borisov, A.B., Longworth, J. W., Boyer, K., and Rhodes, C. K., "Cryptographic Unification of Mass and Space Links Neutrino Flavor ($\nu_e/\nu_\mu$) Transformations with the Cosmological Constant $\Lambda$," *International Journal of Modern Physics* **A18**, 4257 (2003).

2.  Dai, Y., Borisov, A.B., Boyer, K., and Rhodes, C.K., "Computation with Inverse States in a Finite Field $\mathbb{F}_{P_\alpha}$: The Muon Neutrino Mass, the Unified-Strong-Electroweak Coupling Constant, and the Higgs Mass," Sandia National Laboratories, *Report SAND2000-2043*, August 2000.

3.  Lidl, R., and Niederreiter, H., *Finite Fields*. Vol. 20 of the *Encyclopedia of Mathematics and its Applications*, ed. G.-C. Rota (Cambridge University Press, Cambridge, 1977).

4.  Koblitz, N., *A Course in Number Theory and Cryptography* (Springer-Verlag, Berlin/New York, 1987).

5.  Lidl, R. and Niederreiter, H., "Introduction to Finite Fields and their Applications, Revised Edition" (Cambridge University Press, Cambridge, 1994).

6.  Shparlinski, I. E., "Finite Fields: Theory and Computation" (Kluwer Academic Publishers, Dordrecht/Boston/London, 1999).

7.  Rhodes, C.K., "Unique Physically Anchored Cryptographic Theoretical Calculation of the Fine-Structure Constant $\alpha$ Matching both the g/2 and Interferometric High-Precision Measurements," arXiv:gen-ph/1008.4537, August, 2010.

8.  Cohen, H., *Number Theory, Volume 1: Tools and Diophantine Equations* (Springer-Verlag, New York, 2007).

9.  Dai, Y., Borisov, A. B., Longworth, J. W., Boyer, K., and Rhodes, C. K., "Gamma-Ray Bursts and the Particle Mass Scale," *Proceedings of the International Conference on Electromagnetics in Advanced Applications,* ed. R. Graglia (Politecnico di Torino, Torino, 1999) p. 3.

10. Piran, T., Toward Understanding Gamma-Ray Bursts. In *Unresolved Problems in Astrophysics*, (ed. John N. Bahcall and Jeremiah P. Ostriker) (Princeton: Princeton University Press, 1997) p.343.

11. Kulkarni, S.R. et al., "The Afterglow, Redshift and Extreme Energetics of the $\gamma$-Ray Burst of 23 January 1999," *Nature* **398**, 394 (1999).

12. Galama, T.J. et al., "The Effect of Magnetic Fields on $\gamma$-Ray Bursts Inferred from Multi-Wavelength Observations of the Burst of 23 January 1999," *Nature* **398**, 400 (1999).

13. Totani, T., "TeV Bursts of Gamma-Ray Bursts and Ultra-High-Energy Cosmic Rays," *Astrophys. J. Lett.* **509**, L81 (1998).

14. Akerlof, C. et al., "Observations of Contemporaneous Optical Radiation from a $\gamma$-Ray Burst," *Nature* **398**, 394 (1999).

15. Heavens, A., "The Cosmological Model: An Overview and an Outlook," *J. Phys.*: Conference Series 120, 022001 (2008).

16. Carroll, S.M., "The Cosmological Constant," Living Rev. Relativity 3, 1 (2001).

17. Dai, Y., Borisov, A. B., Longworth, J. W., Boyer K., and Rhodes, C. K., "Quadratic Reciprocity, the Higgs Mass and Complexity," *Adv. Stud. Cont. Math.* **10**, 149 (2005).

18. Hanneke, D., Fogwell, J., and Gabrielse, G., "New Measurement of the Electron Magnetic Moment and the Fine Structure Constant," *Phys. Rev. Lett.* **100**, 120801 (2008).

19.    Cadonet, M., de Mirandes, E., Cladé, P., Guellati-Khélifa, S., Schwob, C., Nez, F., Julien, L., and Biraben, F., "Combination of Bloch Oscillations with a Ramsey-Bordé Interferometer: New Determination of the Fine Structure Constant," *Phys. Rev. Lett.* **101**, 230801 (2008).

20.    Mohr, P.J., Taylor, B.N., and Newell, D.B., "CODATA Recommended Values of the Fundamental Constants: 2006," *Rev. Mod. Phys.* **80**, 633 (2008).

21.    Gouvêa, F.Q., *p-Adic Numbers* (Springer-Verlag, Berlin/Heidelberg, 1993).

22.    S. Chandrasekhar, *Newton's Principia for the Common Reader* (Oxford University Press, Oxford, 1995).

23.    Granville, A., "Smooth Numbers: Computational Number Theory and Beyond," Algorithimic Number Theory **44**, 267 (2008).

24.    *Collected Papers of Srinivasa Ramanujan,* edited by G.H. Hardy, P.V. Seshu Aiyar, and B.M. Wilson (AMS Chelsea Publishing, Providence, RI, 1962).

25.    Ramanujan, S., "Highly Composite Numbers," annotated by Jean-Louis Nicolas, The Ramanujan Journal **1**, 119 (1997).

26.    Alauglu, L. and Erdös, P., "On Highly Composite and Similar Numbers," *Trans. Amer. Math. Soc.* **56**, 448 (1944).

27.    Erdös, P., Nicolas, J.-L., and Sárközy, A., "On Large Values of the Divisor Function," *The Ramanujan Journal* <u>2</u>, 225 (1998).

28.    Dai, Y., Borisov, A. B., Boyer, K., and. Rhodes, C. K., "Determination of supersymmetric particle masses and attributes with genetic divisors," Sandia National Laboratories, Report SAND2001-1608, June 2001.

29.    Dai, Y., Borisov, A. B., Boyer, K., and. Rhodes, C. K., "A *p*-adic metric for particle mass scale organization with genetic divisors," Sandia National Laboratories, Report SAND2001-2903, December 2001.

30.    Mullen, G.L., and Mummert, C., *Finite Fields and Applications* (American Mathematical Society, Providence, RI, 2007).

31.    Leedham-Green, C.R., and McKay, S., *The Structure of Groups of Prime Power Order* (Oxford University Press, Oxford, 2002).

32.    Tschebyscheff, P.L., *Theorie der Congruenzen* (Chelsea Publishing Co., New York, 1972).

33.    Hardy, G.H., and Wright, E.M., *An Introduction to the Theory of Numbers*, Fourth Edition (Oxford University Press, Oxford, 1960).

34.    Apostol, T., *Introduction to Analytic Number Theory*, (Springer-Verlag, New York, 1976).

35.    Lemmermeyer, F., *Reciprocity Laws* (Springer-Verlag, Berlin/New York, 2000).

36.    Allen, S.W., Rapetti, D.A., Schmidt, R.W., Ebling, H., Morris, R.G., and Fabian, A.C., "Improved Contraints on Dark Energy from *Chandra* X-ray Observations of the Largest Relaxed Galaxy Clusters," *Mon. Not. R. Astron. Soc.* **383**, 879 (2008).

37.    Christian Marioni and Adeline Bussi, "A Geometric Measure of Dark Energy with Pairs of Galaxies", *Nature* **468**, 539 (2010).

38.    H. Chord, R. Srianand, P. Petitjean, B. Aracil, "Probing the Cosmological Variation of the Fine-Structure Constant: Results Based on VLT-UVES Sample", *Astron. and Astrophys.* **417**, 813 (2004).

39.    R. Quast, D. Reimers, and S.H. Levshakov, "Probing the Variability of the Fine-Structure Constant with the VLT/UVES", *Astron. and Astrophys.* **415**, L7 (2004).